\documentclass[aps,twocolumn,pra,superscriptaddress,amsmath,amssymb]{revtex4-1}
\usepackage{dcolumn}
\usepackage{bm}
\usepackage{amsmath}
\usepackage{txfonts}
\usepackage[T1]{fontenc}
\usepackage{xspace}
\usepackage{ulem}
\usepackage{comment}
\usepackage{braket}
\usepackage{enumerate}
\ifx\pdfoutput\undefined
\usepackage[dvipdfmx]{graphicx}
\usepackage[dvipdfmx]{hyperref}
\usepackage[dvipdfmx]{color}
\else
\usepackage{graphicx}
\usepackage{hyperref}
\usepackage{color}
\fi

\newcommand{\Hzz}{H_{00}}
\newcommand{\Hhh}{H_{\frac{1}{2}\frac{1}{2}}}

\begin{document}
\let\emph\textit

\title{
  Ferromagnetically ordered states in the Hubbard model
  on the $\Hzz$ hexagonal golden-mean tiling 
}

\author{Toranosuke Matsubara}

\affiliation{
  Department of Physics, Tokyo Institute of Technology,
  Meguro, Tokyo 152-8551, Japan
}

\author{Akihisa Koga}

\affiliation{
  Department of Physics, Tokyo Institute of Technology,
  Meguro, Tokyo 152-8551, Japan
}

\author{Sam Coates}

\affiliation{
  Department of Materials Science and Technology,
  Tokyo University of Science, Katsushika, Tokyo 125-8585, Japan
}
\affiliation{Surface Science Research Centre and Department of Physics, University of Liverpool, Liverpool L69 3BX, UK}

\date{\today}
\begin{abstract}
  We study magnetic properties of the half-filled Hubbard model
  on the two-dimensional $\Hzz$ hexagonal golden-mean quasiperiodic tiling.
  The tiling is composed of large and small hexagons, 
  and parallelograms, and its vertex model is bipartite
  with a sublattice imbalance.
  The tight-binding model on the tiling
  has macroscopically degenerate states at $E = 0$.
  We find the existence of two extended states in one of the sublattices,
  in addition to confined states in the other.
  This property is distinct from that of the well-known two-dimensional quasiperiodic tilings
  such as the Penrose and Ammann-Beenker tilings.
  Applying the Lieb theorem to the Hubbard model on the tiling,
  we obtain the exact fraction of the confined states as $1/2\tau^2$, where $\tau$ is the golden mean.
  This leads to a ferromagnetically ordered state in the weak coupling limit.
  Increasing the Coulomb interaction,
  the staggered magnetic moments are induced and gradually increase.
  Crossover behaviour in the magnetically ordered states is also addressed
  in terms of perpendicular space analysis.
\end{abstract}
\maketitle

\section{Introduction}
Quasiperiodic systems have attracted considerable interest 
since the discovery of the Al-Mn quasicrystal~\cite{Shechtman}. Their properties are of equal interest, in part driven by the observation of behaviour traditionally observed in periodic systems. For example, electron correlations in quasicrystals have been 
actively studied after quantum critical behavior was observed 
in the Au-Al-Yb quasicrystal~\cite{Deguchi_2012}.
Similarly, long-range correlative states have been reported despite the lack of periodicity inherent in these materials:
such as superconductivity in the Al-Zn-Mg quasicrystal~\cite{Kamiya_2018}, 
and ferromagnetically ordered states
in the Au-Ga-X (X = Gd, Tb, Dy) quasicrystals~\cite{Tamura, Takeuchi_2023}. 
These studies have stimulated, and continue to motivate theoretical investigations on 
electron correlations and 
the spontaneously symmetry breaking states 
in quasicrystals~\cite{Okabe_1988,Wessel_2003,Jagannathan_2007,Watanabe_2013,Takemori_2015,Takemura_2015,Andrade_2015,Fulga_2016,Otsuki_2016,Sakai_2017,Ara_2019,Sakai_2019,Varjas_2019,Duncan_2020,Cao_2020,Takemori_2020,Hauck_2021,Ghadimi_2021,Sakai_2021}. For example, magnetically ordered states in the Hubbard model 
on quasiperiodic bipartite tilings have been studied, including 
the Penrose~\cite{Jagannathan_2007, Koga_Tsunetsugu}, 
Ammann-Beenker~\cite{Wessel_2003, Jagannathan_Schulz_1997, Koga_AB,Oktel}, and Socolar dodecagonal~\cite{Koga_dodeca}.
One of the common properties among the majority of these studies is the existence of strictly localized states with $E=0$ (i.e.,  confined states)
in the non-interacting case~\cite{KohmotoSutherland,Arai,Koga_Tsunetsugu,Koga_AB,Oktel,Koga_dodeca,Keskiner_2022, Matsubara_conf}.
This leads to
interesting magnetic properties in the weak coupling limit.

Recently, we introduced a family of golden--mean hexagonal and trigonal aperiodic tilings produced using a generalization of de Bruijn's grid method ~\cite{Sam}. In this work, we showcased the structural properties and substitution rules of two `special' cases of this family. These are the $\Hzz$ and $\Hhh$ tilings, where the subscript refers to the tunable grid-shift parameters used in their construction (for more details, see ~\cite{Sam}). These tilings hold distinct structural properties compared to the Penrose, Ammann-Beenker, and Socolar tilings -- not only do they share rotational symmetries associated with periodic systems, but, they also possess a sublattice imbalance due to their vertex structure. However, they are still rooted in the `physical' world of experimentally observed trigonal and hexagonal quasiperiodic systems ~\cite{Woods14,Uri23,Oka21, Coates20}.

It is therefore desirable to study magnetic properties on quasiperiodic systems with sublattice imbalances, in order to systematically understand and compare correlated electron behavior across the widest range of relevant quasiperiodic tilings. In fact, we have already shown the effect which an imbalance has on the magnetic states on one of the special cases from the hexagonal family; the $\Hhh$ hexagonal golden-mean tiling realizes a ferrimagnetically ordered state in the ground state~\cite{Koga_hexagonal}, which is in contrast to that in the Penrose~\cite{Koga_Tsunetsugu}, Ammann-Beenker~\cite{Koga_AB}, and Socolar dodecagonal tilings~\cite{Koga_dodeca} where antiferromagnetically ordered states are realized without a uniform magnetization.

\begin{figure}[htb]
  \begin{center}
    \includegraphics[width=\linewidth]{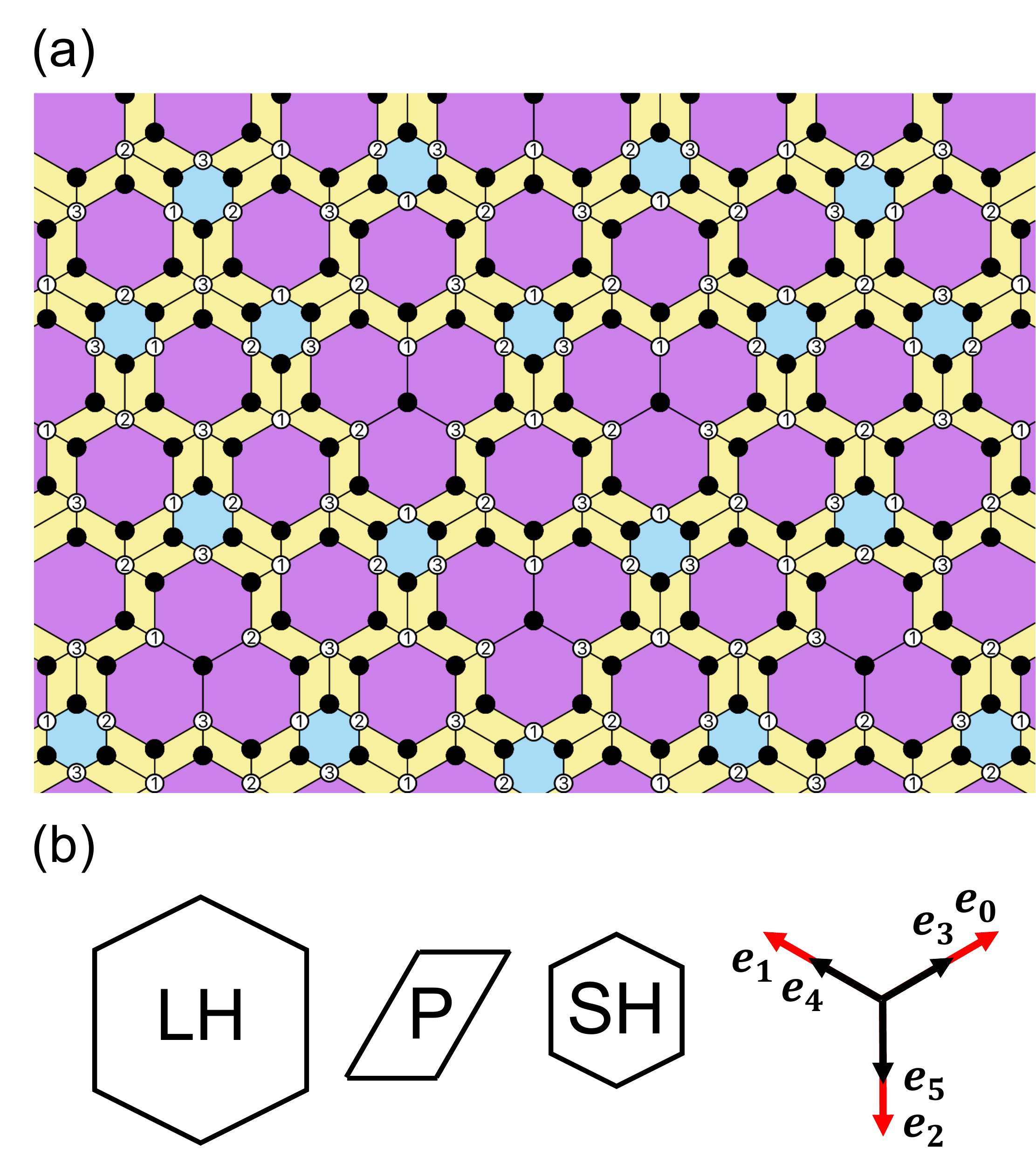}
    \caption{(a) The $\Hzz$ hexagonal golden-mean tiling.
      Open (filled) circles at the vertices indicate the $w$ ($b$) sublattice in the system.
      The numbers in the open circles indicate the indices in the $w$ sublattice (see text).
      (b) Large hexagon, parallelogram, and small hexagon. ${\bm e}_0,\cdots,{\bm e}_5$ are the projection of the fundamental translation vectors in six dimensions, ${\bm n}=(1,0,0,0,0,0),\cdots,(0,0,0,0,0,1)$.}
    \label{fig: Tiling}
  \end{center}
\end{figure}
In this paper, we discuss the relevant properties of the $\Hzz$ tiling structure and then
study the macroscopically degenerate states with $E=0$
in the tight-binding model, which should play an important role for finding magnetic properties 
in the weak coupling limit.
We clarify that two extended states appear in one of the sublattices, 
while confined states appear in the other. 
Furthermore, we obtain the exact fraction of the confined states 
in terms of Lieb's theorem~\cite{Lieb}, 
considering magnetism in the weak coupling limit. 
We also discuss how magnetic properties are affected by electron correlations
in the half-filled Hubbard model.

The paper is organized as follows.
In Sec.~\ref{sec: tiling}, we briefly describe the properties of the $\Hzz$ hexagonal golden-mean tiling needed for our work.
In Sec.~\ref{sec: model}, we introduce the half-filled Hubbard model on the $\Hzz$ hexagonal golden-mean tiling.
Then, we study the macroscopically degenerate states with $E = 0$ 
in Sec.~\ref{sec: conf}.
By means of the real-space Hartree approximations, 
we clarify how a magnetically ordered state is realized 
in the Hubbard model in Sec.~\ref{sec: mag}. 
Finally, crossover behavior in the ordered state is addressed 
by mapping the spatial distribution of the magnetization 
to perpendicular space.
A summary is given in the last section.

\section{Properties of the $\Hzz$ hexagonal golden-mean tiling}
\label{sec: tiling}
Here, we will give an overview and describe the relevant properties of the $\Hzz$ hexagonal golden-mean tiling which we need for our calculations. The tiling is
composed of large hexagons (LH), parallelograms (P),
and small hexagons (SH). A section of the tiling, and the schematics of its proto-tiles are shown in Figure~\ref{fig: Tiling}. The vertex system of the tiling is bipartite, since it is composed of polygons with even edges (hexagons and parallelograms). For our work, we require the exact fractions of tile and vertex frequencies across the tiling, which we take directly from ~\cite{Sam}, in which we explicitly explain our methods of derivation. 

In the thermodynamic limit, the fractions of the LH, P, and SH tiles are given as:
\begin{figure}[htb]
  \begin{center}
    \includegraphics[width=0.9\linewidth]{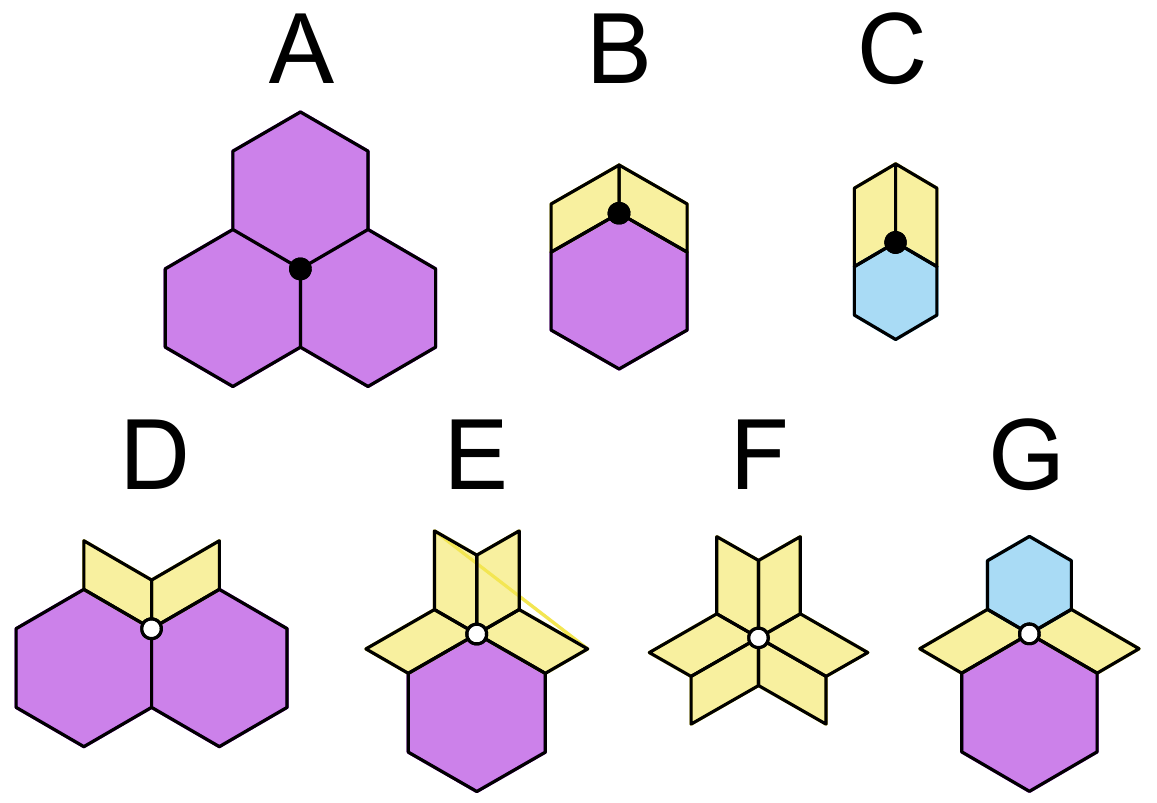}
    \caption{Seven types of vertices.
      Open (filled) circles at the vertices represent $w$ ($b$) sublattice (see text).}
    \label{fig: Deflation_vertex}
  \end{center}
\end{figure}
\begin{eqnarray}
f_{\rm LH} &=& \frac{1}{31} (5\tau -2) \sim 0.196,\\
f_{\rm P} &=& \frac{6}{31} (-2\tau +7) \sim 0.728,\\
f_{\rm SH} &=& \frac{1}{31} (7\tau -9) \sim 0.0750,
\end{eqnarray}
where $\tau$ is the golden-mean $(1+\sqrt{5})/2$.
Similarly, there are seven types of vertices: A, B, C, D, E, F, and G vertices, which are explicitly shown in Figure~\ref{fig: Deflation_vertex}. Their fractions across the tiling are given as:
\begin{eqnarray}
f_{\rm A} &=& \frac{1}{4\tau^5}  \sim 0.0225,\\
f_{\rm B} &=& \frac{3\sqrt{5}}{4\tau^3}  \sim 0.395,\\
f_{\rm C} &=& \frac{3}{4\tau^3}  \sim 0.177,\\
f_{\rm D} &=& \frac{3}{4\tau^5}  \sim 0.0676,\\
f_{\rm E} &=& \frac{3\sqrt{5}}{4\tau^5}  \sim 0.151,\\
f_{\rm F} &=& \frac{1}{4\tau^7} \sim 0.00861,\\
f_{\rm G} &=& \frac{3}{4\tau^3}  \sim 0.177.
\end{eqnarray}

As the tiling is bipartite, trivially, we have two distinct sublattices of vertices. These sublattices can be distinguished either by their occupation of distinct sub-planes in perpendicular space ~\cite{Sam}, or by grouping by their coordination number. For example, one sublattice consists of A, B, and C vertices (coordination number of 3), while the other consists of D, E, F, and G vertices (coordination numbers of 4, 5, 6, and 4, respectively). From here on, the sublattice including A, B, and C vertices is denoted as the $b$ sublattice and the other is denoted as the $w$ sublattice.

As we previously mentioned, this sublattice structure is in contrast to that of the well-known bipartite tilings such as the Penrose and Ammann-Beenker tilings, where half of the vertices for each type exist in both sublattices. The sublattice structure inherent in the $\Hzz$ tiling, however, leads to the sublattice imbalance $\Delta$, such that \cite{Sam}: 
\begin{eqnarray}
\Delta&=&f_{b} - f_{w} = \frac{1}{2\tau^2} \sim 0.190,\\
f_b&=&f_{\rm A}+f_{\rm B}+f_{\rm C} \sim 0.595,\\
f_w&=&f_{\rm D} + f_{\rm E} + f_{\rm F} + f_{\rm G} \sim 0.404,
\label{eq: imbalance}
\end{eqnarray}
where $f_{b}$ and $f_{w}$ are the fractions of the $b$ and $w$ sublattices, respectively.

We note the following property which is convenient for reducing the computational cost of mean-field calculations. In the $\Hzz$ hexagonal tiling, certain tiles or vertices have a local threefold rotational symmetry, {\it e.g.} the LH and SH tiles, and the A and F vertices, as seen in Figure~\ref{fig: Deflation_vertex}. Following the substitution rules in \cite{Sam}, this threefold `group' is changed in a cyclical manner as: LH tile $\rightarrow$ SH tile $\rightarrow$ A vertex $\rightarrow$ F vertex $\rightarrow$ LH tile $\rightarrow$ $\cdots$. Therefore, the system belongs to the point group $C_{3v}$ when one generates the tiling by iteratively applying the deflation rule to an LH or SH tile as its seed, allowing us to save computational time by applying symmetry operations. However, in the thermodynamic limit, the entire system has sixfold rotational symmetry, which is seen in Fourier space~\cite{Sam}.

\section{Model and hamiltonian}
\label{sec: model}
We study the Hubbard model on the $\Hzz$ hexagonal golden-mean tiling,
which is given by the following Hamiltonian
\begin{eqnarray}
H = -t\sum_{(ij),\sigma} (c^\dag_{i\sigma} c_{j\sigma} +{\rm H.c.}) + U\sum_{i} n_{i\uparrow}n_{i\downarrow},
\end{eqnarray}
where $c_{i\sigma}$ $(c^\dag_{i\sigma} )$ annihilates (creates) an electron with spin 
$\sigma \, (=\uparrow ,\, \downarrow)$ at the $i$th site and $n_{i\sigma} = c^\dag_{i\sigma} c_{i\sigma}$. 
$t$ is the nearest-neighbor transfer integral and $U$ is the on-site Coulomb interaction. 
For simplicity, we have assumed that the magnitude of the hopping integral is uniform in the system.
The chemical potential is always $\mu = U/2$
when the electron density is fixed to be half filling.

To discuss magnetic properties in the Hubbard model,
we make use of the real-space mean-field theory.
This method has an advantage in treating large clusters,
which is crucial to clarify magnetic properties inherent in the quasiperiodic systems.
Here, we introduce the site-dependent mean-field $\braket{n_{i\sigma}}$ and
the mean-field Hamiltonian is then given as
\begin{eqnarray}
H^{\rm MF} = -t\sum_{(ij),\sigma} (c^\dag_{i\sigma} c_{j\sigma} +{\rm H.c.}) + U\sum_{i,\sigma} n_{i\sigma} \braket{n_{i\bar{\sigma}}}.
\end{eqnarray}
For given values of $\braket{n_{i\sigma}}$, we numerically diagonalize the Hamiltonian $H^{\rm MF}$, update $\braket{n_{i\sigma}}$, and repeat this procedure until the result converges. The uniform and staggered magnetizations $m^{\pm}$ are given as
\begin{eqnarray}
m^{\pm} &=& f_{b} m_{b} \pm f_{w} m_{w}, \\
m_{\alpha} &=& \frac{1}{N_\alpha} \sum_{i \in \alpha} m_i, \\
m_i &=& \frac{1}{2} \left( \braket{n_{i\uparrow}} - \braket{n_{i\downarrow}} \right),
\end{eqnarray}
where $N_\alpha$ ($m_\alpha$) is the number of the sites (the average of the magnetization) in the $\alpha$ sublattice and
$m_i$ is the local magnetization at the $i$th site.

\begin{figure}[htb]
  \begin{center}
    \includegraphics[width=\linewidth]{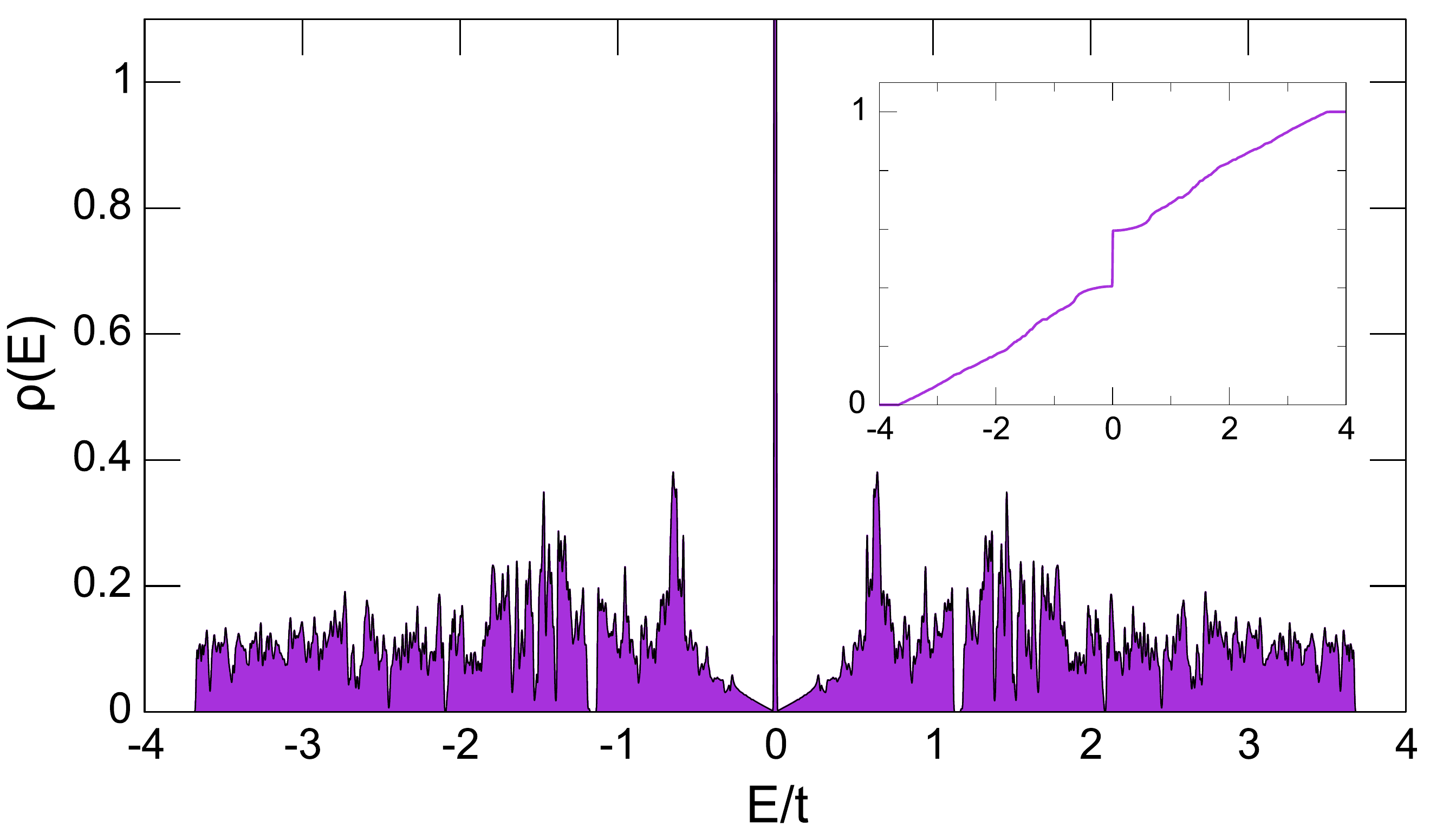}
    \caption{Density of states of the tight-binding model on the $\Hzz$ hexagonal golden-mean tiling with $N = 1\, 767\, 438$. 
    The inset shows the integrated density of states.}
    \label{fig: DOS}
  \end{center}
\end{figure}

Here, we discuss electronic properties in the noninteracting case ($U=0$),
where the model Hamiltonian is reduced to the tightbinding model.
Diagonalizing the Hamiltonian for the system with $N = 1\, 767\, 438$, 
we obtain the density of states as
\begin{eqnarray}
\rho (E) = \frac{1}{N} \sum_i \delta (E - \epsilon_i),
\end{eqnarray}
where $N$ ($=\sum_\alpha N_\alpha$) is the number of the sites in the whole system and $\epsilon_i$ is the $i$th eigenenergy. 
The results are shown in Figure~\ref{fig: DOS}.
We find the delta-function peak at $E=0$, 
suggesting the existence of macroscopically degenerate states. 
In fact, the clear jump singularity appears at $E=0$ in the integrated density of states.
These states should be important for magnetic properties 
in the weak coupling limit. 
In the next section, we discuss the macroscopically degenerate states with $E=0$.

\section{Macroscopically degenerate states} \label{sec: conf}
Here, we focus on the degenerate states with $E=0$ in the tightbinding model.
Since the $\Hzz$ hexagonal golden-mean tiling is bipartite,
these states should exist in both sublattices.
First, we focus on the $w$ sublattice composed of D, E, F, and G vertices.
It is clarified that the number of the degenerate states in the sublattice is at most two,
which will be proven in Appendix~\ref{app: zero in b}. 
This proof is based on the fact that there exist no tiles with zero amplitudes.
Since all tiles have finite amplitudes in either corner site,
this should indicate the existence of {\it extended states}.
To clarify this, we consider the detail of the $w$ sublattice.
Figure~\ref{fig: Tiling}(a) shows that the $w$ sublattice can be divided 
into three groups $(w_1, w_2, w_3)$, which are shown as the numbers in the open circles.
Each site in the $b$ sublattice connects to three nearest-neighbor sites 
belonging to each of the $w_1$, $w_2$, and $w_3$ sublattices.
Again, this is proven in Appendix~\ref{app: B}. 
Therefore, two states $|\Psi_\pm\rangle$ are the exact eigenstates 
with $E=0$ in the $w$ sublattice, where the amplitudes are given as
\begin{equation}
  \begin{array}{lll}
    \displaystyle
  \langle i_1|\Psi_\pm\rangle=1, & \displaystyle\langle i_2|\Psi_\pm\rangle=\omega_\pm,& 
  \displaystyle\langle i_3|\Psi_\pm\rangle=\omega_\pm^2, \\
  \end{array}
  \label{extended}
\end{equation}
where $i_n\;(n=1,2,3)$ is the site index in the $w_n$ sublattice
and $\omega_\pm(=\exp[\pm 2\pi i/3])$ is a solution of 
the equation $x^2+x+1=0$.
Since finite amplitudes appear in the whole system, 
these states can be regarded as the extended states.
Therefore, we can say that there exist only two extended states with $E=0$ 
in the $w$ sublattice.

\begin{figure}[htb]
  \begin{center}
    \includegraphics[width=\linewidth]{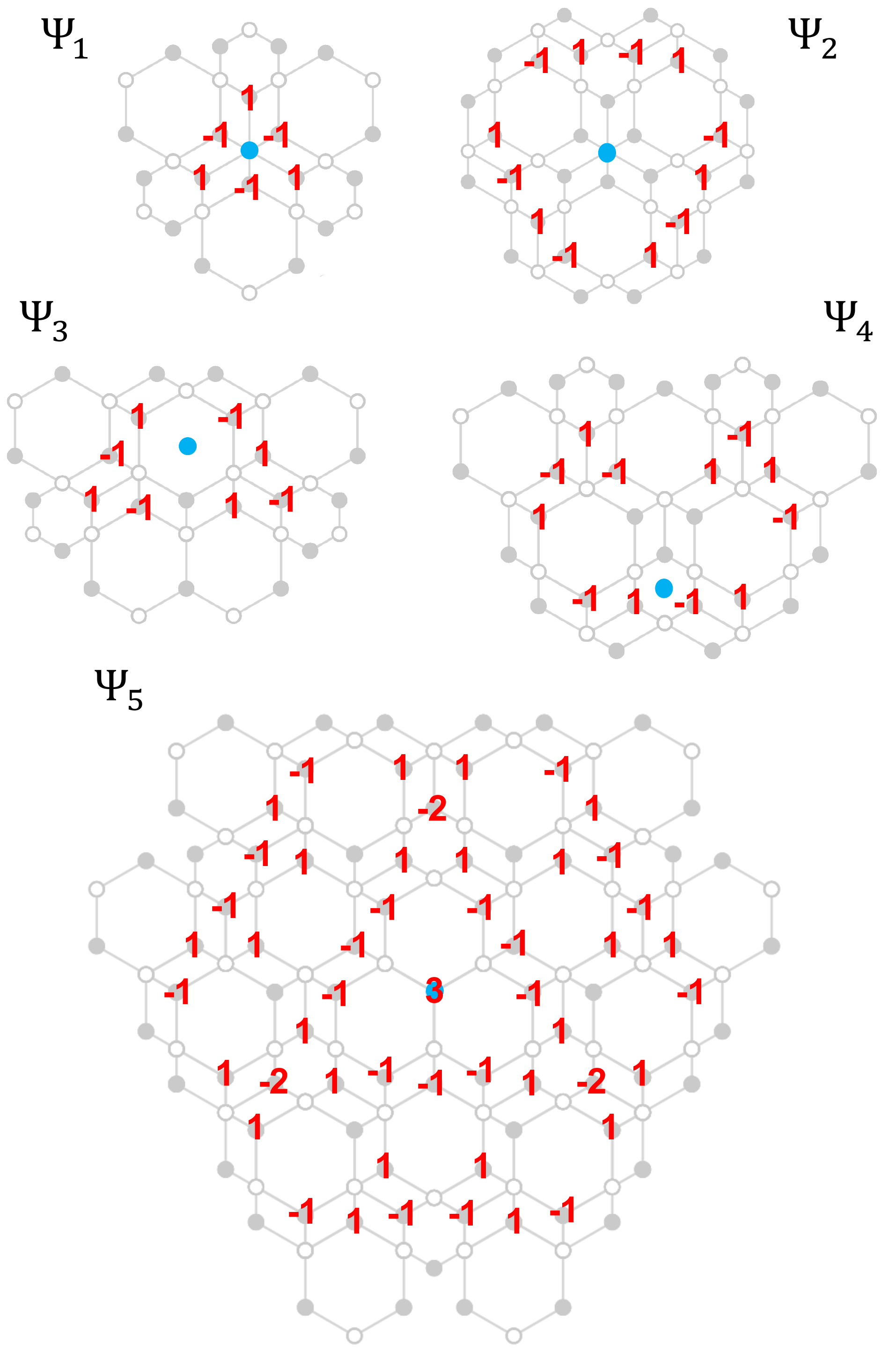}
    \caption{Five confined states in the $b$ sublattice around the vertex or tile with a locally rotational symmetry. 
    Open (filled) circles represent the $w$ ($b$) sublattice. 
    The numbers at the vertices represent the amplitudes of the confined states.}
    \label{fig: conf}
  \end{center}
\end{figure}

By contrast, there are the other macroscopically degenerate states in the $b$ sublattice.
We construct a simple form, considering their linear combinations,
as discussed in previous papers~\cite{Koga_Tsunetsugu, KohmotoSutherland, Arai}.
The states can be represented to be exactly localized in a certain region
and can be regarded as {\it confined states}. 
These are in contrast to the extended states in the $w$ sublattice.
Five simple examples of the confined states $\Psi_1, \Psi_2, \cdots,$ and $\Psi_5$
are explicitly shown in Figure~\ref{fig: conf}.
According to Conway's theorem,
a certain diagram appears repeatedly in the quasiperiodic tiling, in general.
This means that each confined state exists with a finite fraction in the tiling. 
The diagram for the site structures of $\Psi_1$ and $\Psi_2$, which is shown in Figure~\ref{fig: conf}, 
always appears around the F vertex due to the matching rule of the tiles.
Therefore, the fractions for these confined states are given by
the fraction of the F vertex, $f_1=f_2=1/(4\tau^7)$.
On the other hand, the site structures for $\Psi_3$, $\Psi_4$, and $\Psi_5$ 
do not always appear around the LH and SH tiles, and A vertices, respectively.
Taking the tiling structure into account, we obtain the fractions of $\Psi_3 $, $\Psi_4$, and $\Psi_5$ 
as $f_3=(\tau^{-3}+\tau^{-8})/4$, $f_4=3/4\tau^7$, and $f_5=(\tau^{-6}-\tau^{-11})/4$, respectively.
In the tightbinding model on the $\Hzz$ hexagonal golden-mean tiling, 
there are many kinds of confined states (not shown)
and therefore the fraction of the confined states $f^C$ is
bounded by $f^C\ge\sum_{i=1}^5 f_i \sim 0.120$. 

Next, we try to directly obtain the exact fraction of the confined states, 
making use of magnetic properties at half filling~\cite{Koga_hexagonal}.
According to Lieb's theorem,
the ground state of the half-filled Hubbard model has a total spin 
$S_{tot}=N\Delta/2=N/(4\tau^2)$ for arbitrary $U$.
In the weak coupling limit, 
the magnetically ordered state originates only from
the macroscopically degenerate states with $E=0$.
Two extended states in the $w$ sublattice 
should be negligible in the thermodynamic limit. 
Therefore, magnetic properties little depend on these states
and mainly depend on the confined states in the $b$ sublattice.
Thus, the uniform magnetization can be given as $m^+ = f^C/2$, 
where $f^C$ is the fraction of the confined states. 
From these two equations, we obtain the exact fraction of the confined states as
\begin{eqnarray}
f^C = \frac{1}{2\tau^2}  \sim 0.190.
\end{eqnarray}
This is consistent with the numerical results 
$f^C=336288/1767438 \sim 0.190$ for the finite cluster with $N=1\, 767\, 438$.

We wish to note that in the $\Hzz$ hexagonal golden-mean tiling
the extended states appear in addition to the confined states. 
The extended states are also found in the tight-binding model on the $\Hhh$ hexagonal tiling, 
although this was not discussed in our previous work~\cite{Koga_hexagonal}.

\section{Magnetic properties}
\label{sec: mag}
\begin{figure}[htb]
  \begin{center}
    \includegraphics[width=0.8\linewidth]{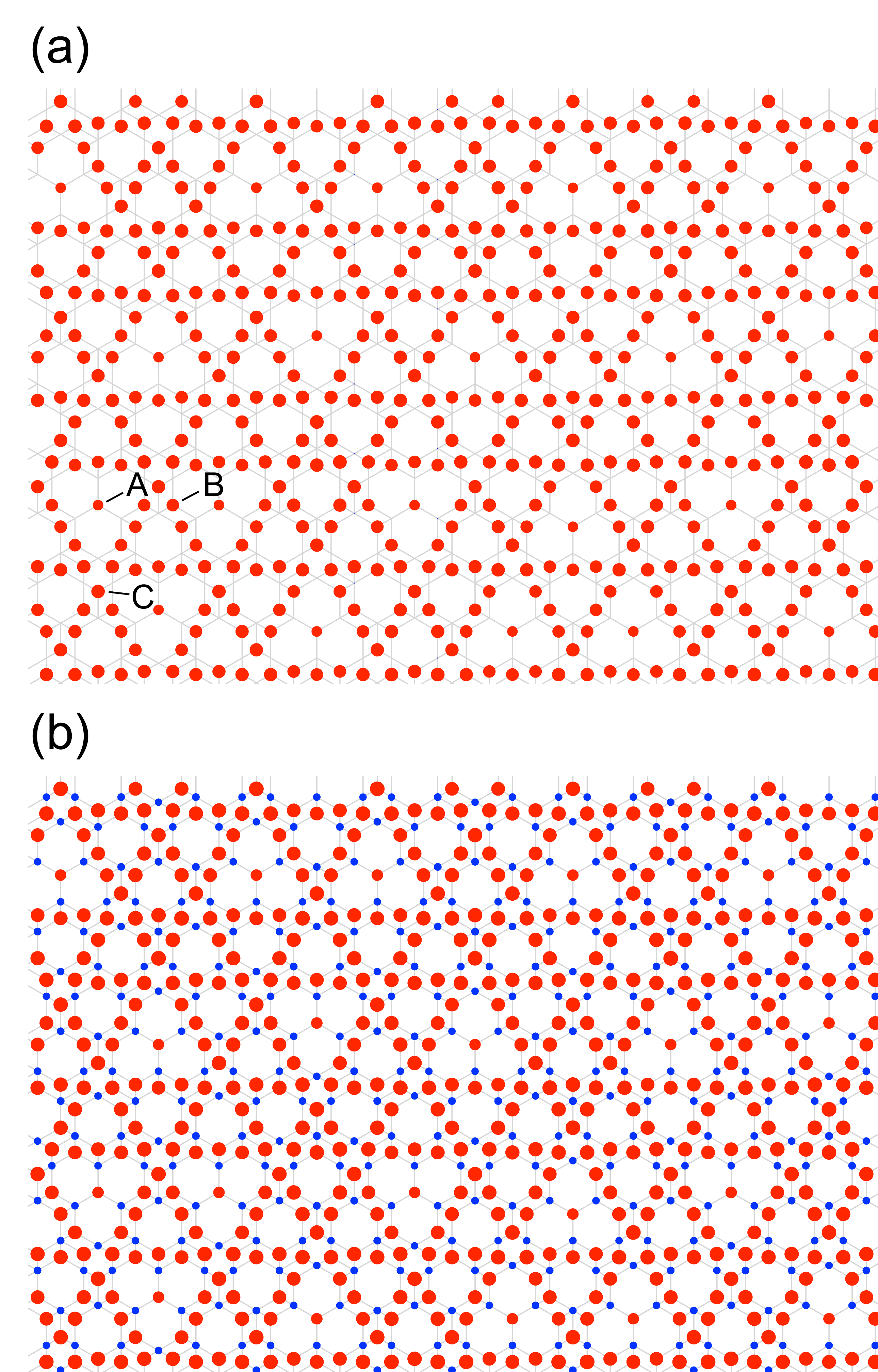}
    \caption{Spatial pattern for the magnetizations in the Hubbard model on the $\Hzz$ hexagonal golden-mean tiling when (a) $U/t = 1.0 \times 10^{-7}$ (essentially the same as $U/t \rightarrow 0$) and (b) $U/t = 1$. The area of the circles represents the magnitude of the local magnetization. Red (blue) filling represents positive (negative) sign. 
    }
    \label{fig: mag_RS}
  \end{center}
\end{figure}

\begin{figure}[htb]
  \begin{center}
    \includegraphics[width=0.8\linewidth]{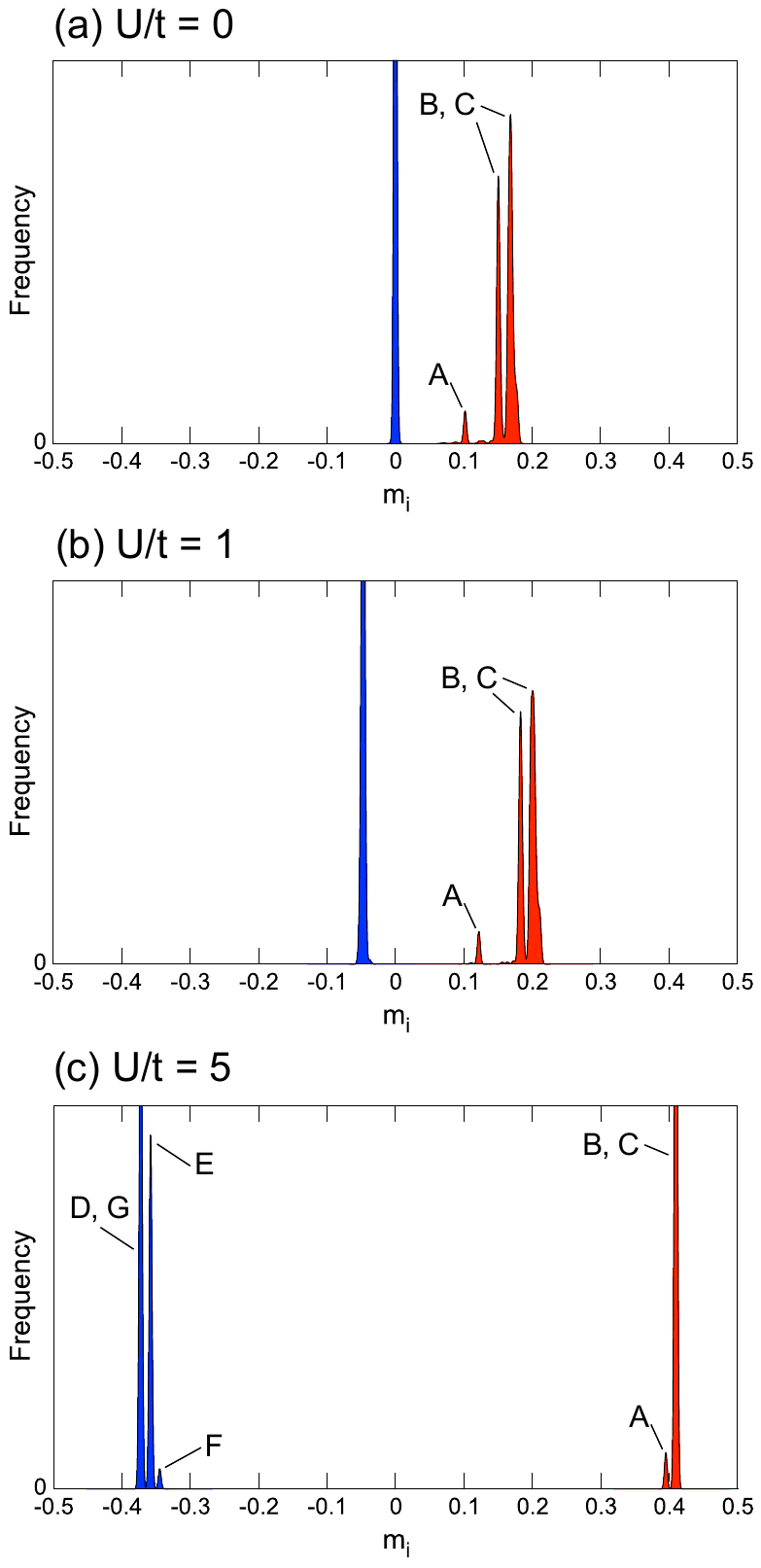}
    \caption{Distribution of the magnetizations in the Hubbard model
    with $N=256\, 636$ when (a) $U/t = 1.0 \times 10^{-7}$ (essentially the same as $U/t \rightarrow 0$), (b) $U/t=1$, and (c) $U/t=5$.
    Red (blue) filling represents $m_i$ in $b$ ($w$) sublattice.
}
    \label{fig: fre}
  \end{center}
\end{figure}

Here, we discuss magnetic properties in the half-filled Hubbard model 
on the $\Hzz$ hexagonal golden-mean tiling. 
We mainly treat the system with $N = 256\, 636$ by means of real-space mean-field approximations. 
When the system is non-interacting, the macroscopically degenerate states appear at the Fermi level, 
as shown in Figure~\ref{fig: DOS}. 
The introduction of interaction leads to a magnetically ordered state with finite magnetizations: the magnetization profile for the case with $U/t = 1.0 \times 10^{-7}$ is shown in Figure~\ref{fig: mag_RS}(a), 
where red circles indicate positive magnetizations, and its size is proportional to the magnitude. 
We find finite magnetizations only in the $b$ sublattice, as discussed above.
In particular, the magnetizations in the A vertices are smaller than 
those in the B and C vertices.
This quantitative difference is clearly found in the distribution of the magnetization in Figure~\ref{fig: fre}(a), 
where the magnetizations on the A vertices are $m\sim 0.1$, while those on the B and C vertices are $m\sim 0.16$ . 

This behaviour can be explained by the spatial distribution of the confined states.
When one considers the local tiling structure for the confined states 
$\Psi_1, \Psi_2, \cdots, \Psi_5$ (see Figure~\ref{fig: conf}),
the A vertices have an amplitude only in the wave function $\Psi_5$.
On the other hand, multiple confined states have amplitudes at B and C vertices.
This should lead to a difference in the magnetizations, namely, the confined states in the larger regions have amplitudes in many sites
and therefore have a minimal effect on the magnetizations at the A vertices.
By contrast, we find no magnetization in the $w$ sublattice, 
which is consistent with the fact that 
two extended states little affect magnetic properties in the weak coupling limit.  
From these results, we can say that, in the weak coupling limit, the ferromagnetically ordered state 
is realized with the total uniform moment $m^+=1/(4\tau^2)$.

Increasing the interaction strength, the local magnetization in the $b$ sublattice monotonically increases 
and the magnetizations in the other sublattice are induced.
The spatial structure in the magnetization for $U/t=1$ is shown in Figure~\ref{fig: mag_RS}(b). 
The magnetization $m\sim -0.05$ is induced in the $w$ sublattice, as shown in Figure~\ref{fig: fre}(b).
Further increasing the interaction strength $U$ changes the distribution of the local magnetizations: when $U/t=5$, the magnetization is almost $m\sim \pm 0.4$, as shown in Figure~\ref{fig: fre}(c).

\begin{figure}[htb]
  \begin{center}
    \includegraphics[width=\linewidth]{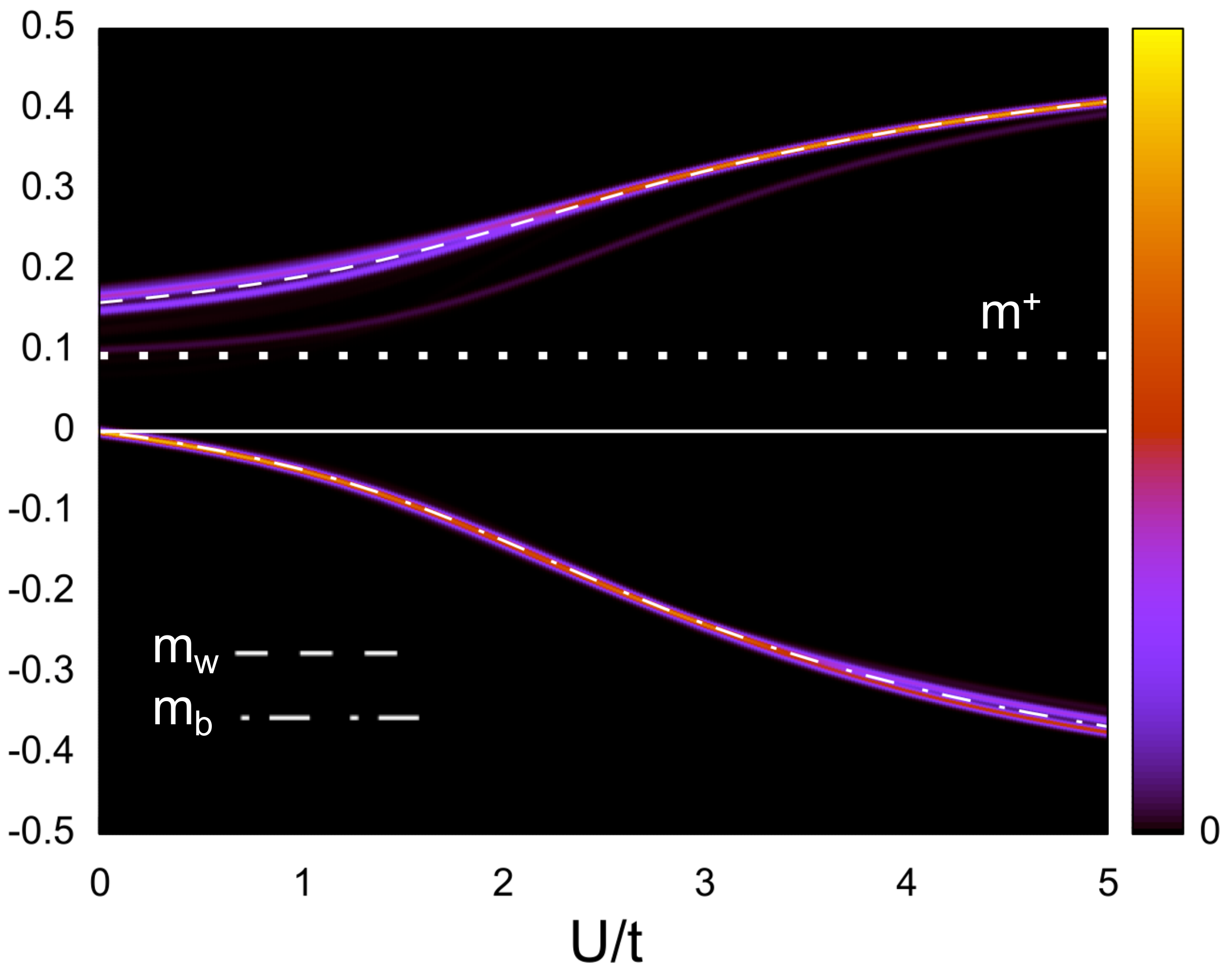}
    \caption{Distribution of the local magnetizations as a function of $U/t$ in the system with $N =97\, 560$. The dashed (dot-dashed) line represents the magnetization $m_b$ ($m_w$), and the dotted line represents the total uniform magnetization.}
    \label{fig: U_m}
  \end{center}
\end{figure}

Figure~\ref{fig: U_m} shows the change in the distribution of the local moments. When $U/t \lesssim 1$, the distribution is similar to that in the weak coupling limit $U/t \rightarrow +0$. 
Namely, a sharp peak appears at $m<0$ (the $w$ sublattice), 
while some peaks appear at $m > 0$ (the $b$ sublattice).
When $U/t\gtrsim 3$, distinct behavior appears in the magnetic distribution. 
In the strong coupling case, the local magnetization should be classified into some groups. 
In the $b$ sublattice with $m>0$, 
the magnetization at the A vertices is distinct from that at the B and C vertices,
and this behavior appears in the whole parameter space.
On the other hand, in the $w$ sublattice with $m<0$, 
the magnetization is classified into three groups characteristic of the coordination number $z$.
Namely, $m\sim -0.37$ for D and G vertices with $z=4$, $m\sim -0.36$ for the E vertices, 
and $m\sim -0.35$ for the F vertices when $U/t=5$, as shown in Figure~\ref{fig: fre}(c).
This is distinct from the weak coupling case.

The crossover between the weak and strong coupling regimes occurs around $U/t \sim 2$.
In the strong coupling limit $U/t \rightarrow \infty$, 
the Hubbard model is reduced to the antiferromagnetic Heisenberg model with nearest-neighbor couplings $J=4t^2/U$. 
The mean-field ground state is described by the staggered moment $m_j \rightarrow \pm 1/2$. 
This means that the mean-field approach cannot correctly describe 
the reduction of the magnetic moment due to quantum fluctuations. 
Therefore, an alternative method is necessary to clarify magnetic properties in this regime, 
which is beyond the scope of the present study. 
Nevertheless, interesting magnetic properties inherent in 
the $\Hzz$ hexagonal golden-mean tiling,
{\it eg.} the ferromagnetically ordered state in the weak coupling limit,
can be captured even in our simple mean-field method.

\begin{figure}[htb]
  \begin{center}
    \includegraphics[width=\linewidth]{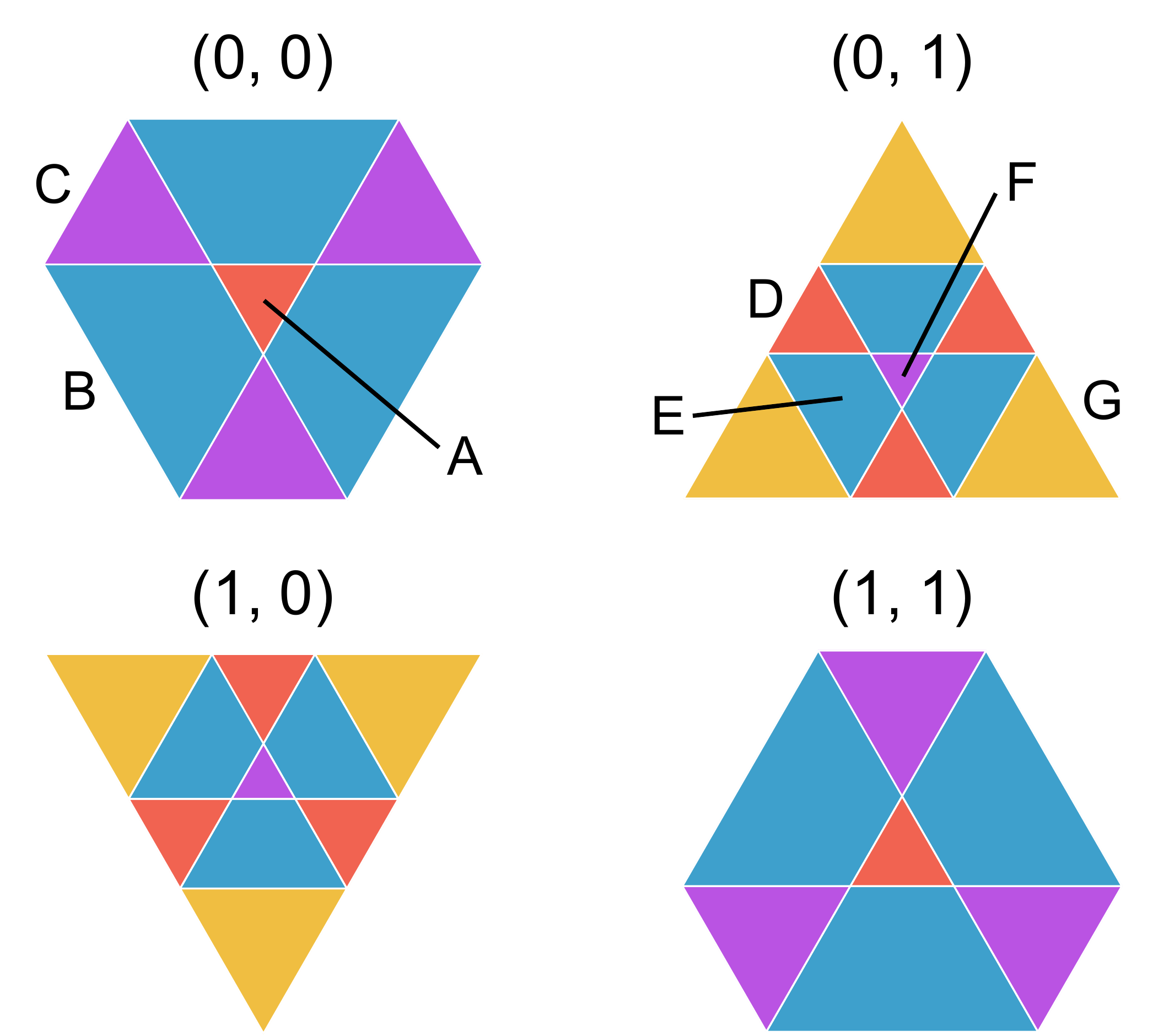}
    \caption{Perpendicular spaces ${\bm r}^\perp$ for the $\Hzz$ hexagonal golden-mean tiling. 
    Each area bounded by the solid lines is the region of one of 7 types of vertices 
    shown in Figure~\ref{fig: Deflation_vertex}.}
    \label{fig: PS}
  \end{center}
\end{figure}

Finally, we wish to demonstrate the spatial profile of the magnetizations characteristic of 
the $\Hzz$ hexagonal golden-mean tiling. 
To this purpose, we map the tiling to perpendicular space ${\bm r}^\perp$, 
where the positions in perpendicular space have one-to-one correspondence with the positions in physical space. We have previously shown that there are four ${\bm r}^\perp$ windows, which can be labelled by pairs of integer heights ~\cite{Sam}, which we re-label here. These heights correspond to where each vertex of the tiling projects onto the body-diagonals of the two 3--dimensional cubes which can be formed by the 6--dimensional super-space basis vectors ${\bm n} = (n_0, n_1, \cdots , n_5)$ ~\cite{Sam}. Thus, the four ${\bm r}^\perp$ planes of the $\Hzz$ hexagonal golden-mean tiling are described by these heights as ${\bm r}^\perp = (x^\perp , y^\perp)$ where $x^\perp = 0, 1$ and $y^\perp = 0, 1$. The A, B, and C vertices uniquely occupy the (0, 0) and (1, 1) planes, with the remaining vertices occupying the remaining planes, which is schematically shown in Figure~\ref{fig: PS}.

\begin{figure*}[htb]
  \begin{center}
    \includegraphics[width=14.0cm]{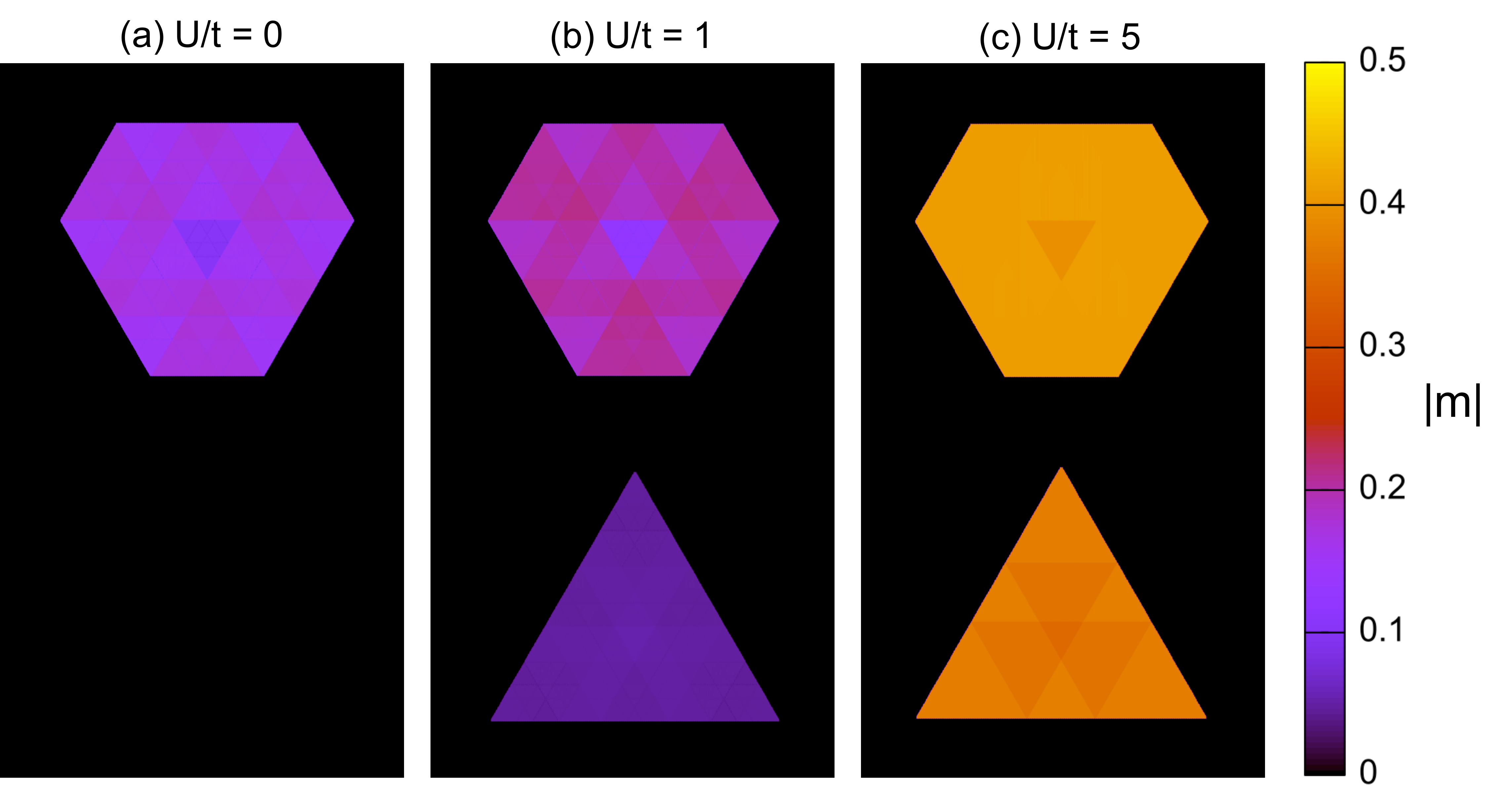}
    \caption{Magnetization profile in the perpendicular space
    for the Hubbard model with $N = 256\, 636$ when (a) $U/t = 1.0 \times 10^{-7}$, (b) $U/t =1$, and (c) $U/t =5$.}
    \label{fig: mag_PS}
  \end{center}
\end{figure*}
The magnetization profiles of the (0, 0) and (0, 1) planes in perpendicular space are shown in Figure~\ref{fig: mag_PS}, 
where we show the absolute values of the local magnetizations. As the (1, 0) and (1, 1) planes are equivalent, it is unnecessary to show them.
When $U/t = 1.0 \times 10^{-7}$, 
the system is essentially the same as that with $U\rightarrow 0$, where
no magnetization appears in the planes (0, 1) and (1, 0) for the $w$ sublattice.
This is consistent with the fact that 
the extended states have little effect on the magnetic properties in the $w$ sublattice.
By contrast, 
finite magnetization appears across the entirety of the (0, 0) and (1, 1) planes, 
implying that the spontaneous magnetizations appear in the $b$ sublattice.
Therefore, we can say that the ferromagnetically ordered state is realized in the weak coupling limit.

\begin{figure}[htb]
  \begin{center}
    \includegraphics[width=0.8\linewidth]{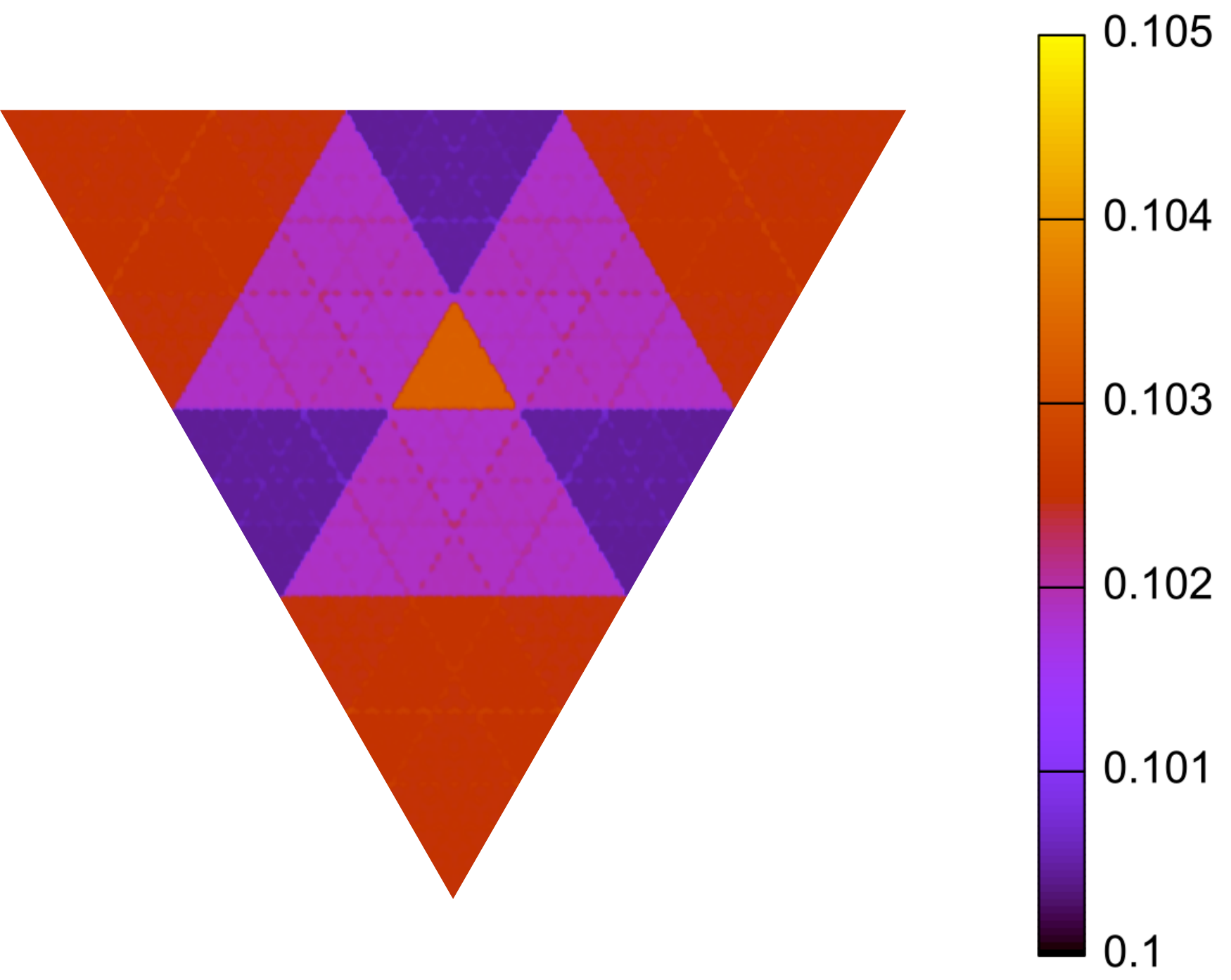}
    \caption{Magnetization profile for the A vertices in the perpendicular space 
    for the Hubbard model with $N = 673\, 873$ when $U/t = 1.0 \times 10^{-7}$ (essentially the same as $U\rightarrow 0$).}
    \label{fig: A}
  \end{center}
\end{figure}
We also find a spatial pattern in the B and C vertex regions in the (0, 0) and (1, 1) planes, and, a spatial pattern with a tiny difference appears in the magnetic moments in the A region,
as shown in Figure~\ref{fig: A}. These suggest the existence of many kinds of confined states in relatively large regions.
This is because the overlapping structure in the confined states should
classify the vertices into hierarchical groups, which yields 
a detailed structure in perpendicular space,
distinct from the simple pattern for the vertices (see Figure~\ref{fig: PS}).

Upon increasing the interaction strength, all vertex sites have magnetizations, as shown in Figure~\ref{fig: mag_PS}(b). 
In the strong coupling case, the Coulomb interactions become crucial  
to stabilize the ferrimagnetically ordered states with staggered moments.
When $U/t = 5$, the local magnetization takes large values. 
In this case, the magnitude of local magnetizations can be classified into two groups in the $b$ sublattice
and three groups in the $w$ sublattice, discussed above.

Before concluding, we would like to summarize and compare the magnetic properties in the Hubbard models on 
the Penrose, Ammann-Beenker, Socolar dodecagonal, $\Hhh$ hexagonal golden-mean, 
and $\Hzz$ hexagonal golden-mean tilings. 
One of the common features is the existence of confined states at $E = 0$ in the noninteracting case ($U = 0$), 
which play a crucial role in stabilizing the magnetically ordered states in the weak coupling limit. 
Nevertheless, their confined state properties are distinct from each other.
The number of types of confined states are six in the Penrose case~\cite{KohmotoSutherland, Arai}, 
while it should be infinite in the others.
As for sublattice structures,
the $\Hhh$ hexagonal golden-mean tiling has a sublattice imbalance, 
leading to a ferrimagnetically ordered state even in the weak coupling limit~\cite{Koga_hexagonal}.
In our $\Hzz$ tiling, however, there exists a sublattice imbalance such that the confined states appear in one of the sublattices,
leading to a ferromagnetically ordered state in the weak coupling limit. 
The other tilings have an equivalent sublattice structure, 
and the corresponding Hubbard model shows the antiferromagnetically ordered state
without a uniform magnetization.

\section{Summary}
\label{sec: summary}
We have studied magnetic properties in the half-filled Hubbard model on the $\Hzz$ hexagonal golden-mean tiling
by means of the real-space mean-field approach. We have found the delta-function peak in the density of states of the tight-binding model, 
implying the existence of macroscopically degenerate confined states at $E = 0$.
We have then clarified that two extended states exist in the $w$ sublattice and 
the confined states appear only in the $b$ sublattice.
For the above properties, we have obtained the exact fraction of the confined states as $1/2\tau^2$.
The introduction of the Coulomb interaction lifts the macroscopic degeneracy at the Fermi level and 
drives the system to a ferromagnetically ordered state.
We have clarified how the spatial distribution of the magnetizations continuously changes with increasing interaction strength. 
Crossover behaviour in the magnetically ordered states has been discussed
by applying perpendicular space analysis to the local magnetizations.

\begin{acknowledgments}
We would like to thank T. Dotera for valuable discussions.
Parts of the numerical calculations were performed 
on the supercomputing systems at ISSP, The University of Tokyo.
This work was supported by Grant-in-Aid for Scientific Research from
JSPS, KAKENHI Grant Nos.
JP22K03525, JP21H01025, JP19H05821 (A.K.), and the EPSRC grant EP/X011984/1 (S.C.).

\end{acknowledgments}

\appendix
\section{Upper bound of the number of the states with $E=0$ in the $w$ sublattice}
\label{app: zero in b}

We examine the number of the states with $E=0$ in the $w$ sublattice
for the noninteracting Hamiltonian $H_0$. 
The states with $E=0$ in the $w$ sublattice can be described as follows,
\begin{eqnarray}
\ket{\Psi} = \sum_{i \in b} \Psi_i \ket{i},
\end{eqnarray}
where $\ket{i}$ is the local state at the $i$th site and $\Psi_i$ is its coefficient. 
The equation $H_0\ket{\Psi}=0$ is reduced to the following simultaneous equation,
\begin{eqnarray}
\sum_{(ij)} \Psi_i = 0,\label{eq:zero}
\end{eqnarray}
where $j$ is the site index in the $b$ sublattice and 
the summation runs to the nearest-neighbor sites of the $i$th site.
The number of the equations is given by $N_b$ and
the number of coefficients is given by $N_w$.
Although $N_b>N_w$, the solutions of eq.~(\ref{eq:zero}) and their number
should not be trivial due to the quasiperiodic structure in the tiling.

To clarify the upper bound of the number of solutions,
we consider a certain domain which is composed of finite tiles connected by the shared edges.
Then, we define a "forbidden domain" so that $\Psi_i=0$ for the vertices inside and on its boundary.
By taking the matching rule of tiles into account,
we sometimes find that, on a certain tile outside of the forbidden domain and adjacent to its boundary,
the amplitudes of the vertices are zero.
This allows us to redefine the forbidden domain to include the tile.
In the other words, the forbidden domain can be regarded as to be {\it expanded}.
In the following, we demonstrate that 
the forbidden domain can be expanded to the whole system
and clarify the upper bound of the number of the degenerate states with $E=0$.

\begin{figure}[htb]
  \begin{center}
    \includegraphics[width=5.7cm]{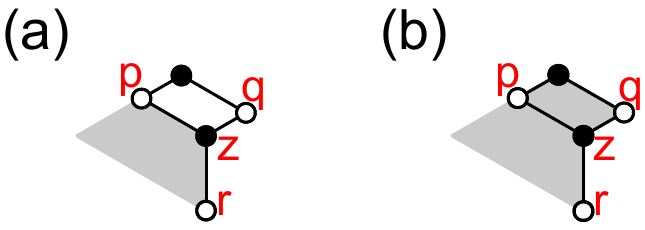}
    \caption{(a) The P tile adjacent to the forbidden domain (shaded area). 
    (b) By taking into account the equation (\ref{eq:zero}), 
    the forbidden domain is expanded. See text.}
    \label{fig: string_P}
  \end{center}
\end{figure}
First, we focus on a P tile outside of the forbidden domain and adjacent to its boundary, 
as shown in Figure~\ref{fig: string_P}(a).
Here, we have labeled three sites in the $w$ sublattice as $p$, $q$, and $r$.
The site $p$ is located on the shared edge, and the site $q$ is located on the other corner of the P tile.
The site $z$ on the shared edge in the $b$ sublattice connects to the nearest-neighbor sites $p$, $q$, and $r$. Since the definition of the forbidden domain, $\Psi_p=0$, and the site $r$ must be on the boundary of the forbidden domain, we obtain $\Psi_r=0$.
We then obtain $\Psi_q=0$ since $\Psi_p+\Psi_q+\Psi_r=0$ [eq.~(\ref{eq:zero})].
Therefore, each site on the P tile has no amplitude, meaning that 
the forbidden domain is expanded to include the P tile, as shown in Figure~\ref{fig: string_P}(b).
By taking into account the above rule, 
the forbidden domain can be expanded so that no P tiles touch outside it.
In the $\Hzz$ hexagonal golden-mean tiling, 
the P tiles densely exist with their fraction $f_P\sim 0.728$
and some of them are connected to each other (see Figure~\ref{fig: Deflation_vertex}).
Therefore, the forbidden domain should be expanded according to the above rules.

\begin{figure}[htb]
  \begin{center}
    \includegraphics[width=7cm]{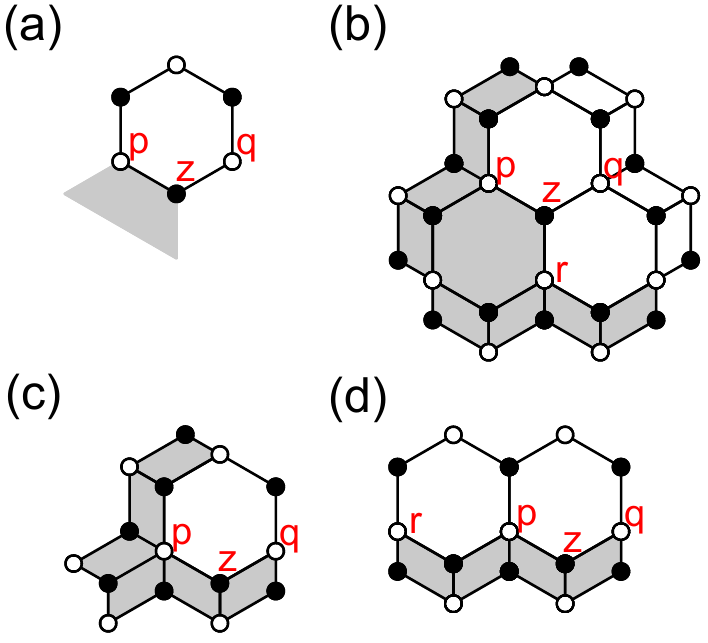}
    \caption{
    (a) The LH tile adjacent to the forbidden domain (shaded area). 
    Two sites on the shared edge are denoted as $p$ and $z$.
    (b) The tiling structure when the A vertex sits on the site $z$.
    (c) The tiling structure when the B vertex sits on the site $z$ and the E vertex sits on the site $p$.
    (d) The tiling structure when the B vertex sits on the site $z$ and the D vertex sits on the site $p$.
}
    \label{fig: string_L}
  \end{center}
\end{figure}

Next, we focus on a certain LH tile outside of the forbidden domain and adjacent to its boundary,
as shown in Figure~\ref{fig: string_L}(a).
We have assumed that the LH tile and forbidden domain share the edge with the sites $p$ and $z$, 
which belong to the $w$ and $b$ sublattices, respectively.
When the A vertex is located at the site $z$, 
the local tiling structure is shown in Figure~\ref{fig: string_L}(b).
The LH tile in the forbidden domain is adjacent to four P tiles and
some P tiles are also connected to each other.
Therefore, the forbidden domain should be expanded, 
which is shown as the shaded area in Figure~\ref{fig: string_L}(b).  
Furthermore, $\Psi_p+\Psi_q+\Psi_r=0$ according to eq.~(\ref{eq:zero}).
Therefore, we conclude that the amplitudes of all corner sites of three LH tiles are zero
and the forbidden domain is expanded to include two LH tiles.

\begin{figure}[htb]
  \begin{center}
    \includegraphics[width=5.7cm]{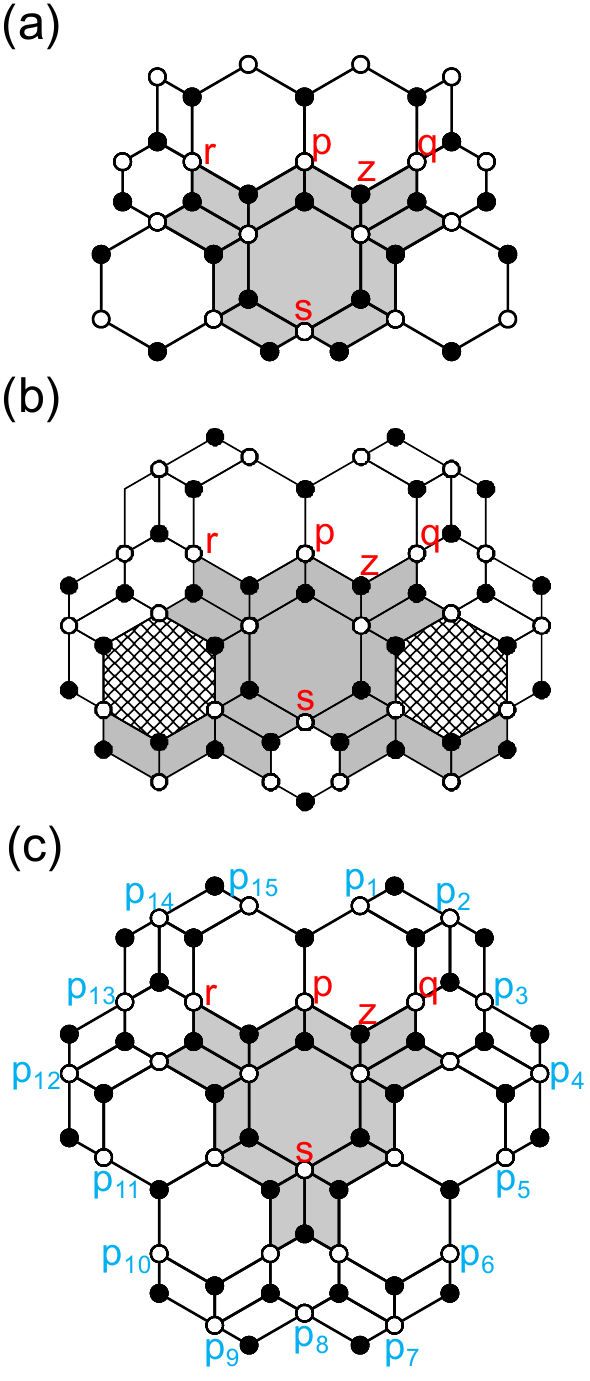}
    \caption{
    The tiling structures when the B vertex sits on the site $z$, the D vertex sits on the site $p$.
    (a) The tiling structure when the G vertices sit on both sites $r$ and $q$.
    (b) The tiling structure when the G vertex sits on both sites $r, q$, and $s$.
    (c) The tiling structure when the G vertices sit on both sites $r$ and $q$, 
    and the E vertex sits on the site $s$.
    The sites marked $p_1, p_2, \cdots, p_{15}$ 
    are used for the proof (see text).
    }
    \label{fig: FD_D}
  \end{center}
\end{figure}

\begin{figure}[htb]
  \begin{center}
    \includegraphics[width=8.5cm]{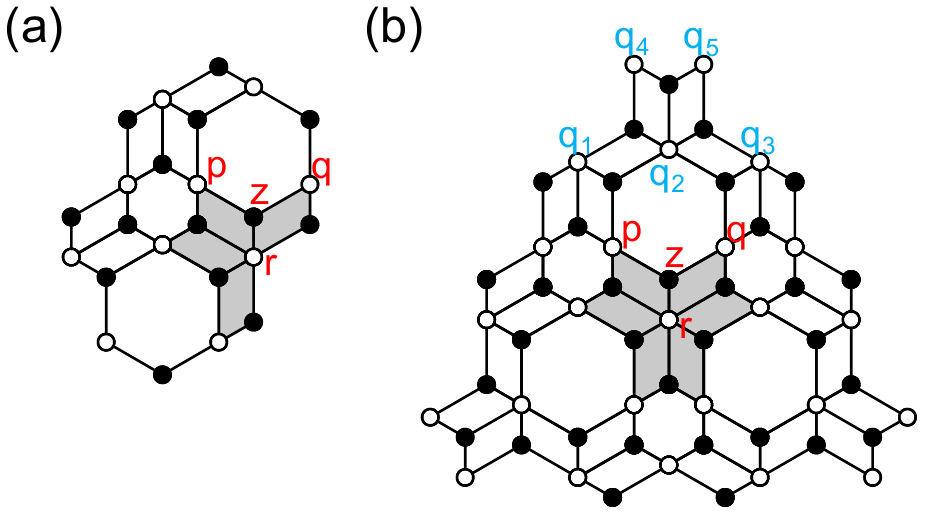}
    \caption{
    (a) The tiling structure when the G vertex sits on the site $p$.
    (b) The tiling structure when the G vertices sit on both sites $p$ and $q$.
    The sites marked $q_1, q_2, \cdots, q_5$ 
    are used for the proof (see text).
        }
    \label{fig: FD_G}
  \end{center}
\end{figure}

When the B vertex is located at the site $z$, 
it is necessary to consider three cases according to the type of vertex at the site $p$; D, E, and G vertices:
\begin{enumerate}[(i)]
	\item the E vertex is located at the site $p$:
	the local structure around the site $p$ is shown in Figure~\ref{fig: string_L}(c).
	The amplitudes of the corner sites of the LH tile are zero 
	since three sites belonging to the $w$ sublattice share the connected P tiles.
	Therefore, the forbidden domain can be spatially expanded to include the LH tile.
	\item the D vertex is located at the site $p$:
	the local structure is symmetric, as shown in Figure~\ref{fig: string_L}(d).
	At the sites $q$ and $r$ the D vertex can not be found due to the matching rule of tiles, however, E or G vertices can be.
	In the case of the E vertex being located at either $q$ or $r$ then we essentially find the same case as (i) and 
	thereby the forbidden domain is expanded to include these two LH tiles.
	When G vertices are located at \textit{both} sites $q$ and $r$,
	the local structure is shown in Figure~\ref{fig: FD_D}(a).
	In this case, either E or G vertex is located at the site $s$. 
	When it is the G vertex, the local structure is shown in Figure~\ref{fig: FD_D}(b). 
	In this case, the forbidden domain is expanded to include the LH tiles,
	which are shown as the hatched hexagons, and the P tiles adjacent to them.
	Furthermore, the forbidden domain is expanded to include the S tiles sharing the sites $q$ and $r$.
	Therefore, finally, the forbidden domain is expanded to include all the tiles shown in Figure~\ref{fig: FD_D}(b).
	On the other hand, when the E vertex is located at the site $s$,
	the vertex structure is shown in Figure~\ref{fig: FD_D}(c). 
	In this case, one may not expand the forbidden domain, 
	which is shown as the shaded area, to a larger domain in terms of our simple rule.
	Now, we must consider the fifteen vertices in the $w$ sublattice outside the area,  
	which is denoted as $p_1$, $p_2$, $\cdots$, $p_{15}$.
	The equations (\ref{eq:zero}) are explicitly given as
	\begin{eqnarray}
		\Psi_{p_1}+\Psi_{p_2}&=&0,\\
		\Psi_{p_2}+\Psi_{p_3}&=&0,\\
		&\vdots&\nonumber\\
		\Psi_{p_{15}}+\Psi_{p_1}&=&0,
	\end{eqnarray}
	since the amplitudes of the vertices on the shaded area are zero. 
	Thus, we obtain $\Psi_{p_1}=\Psi_{p_2}=\cdots=\Psi_{p_{15}}=0$.
	This means that the amplitudes of the vertices on the LH and SH tiles outside of the area must be zero and 
	the forbidden domain can be expanded to include these tiles.
	
	\item the G vertex is located at the site $p$:
	the local structure is shown in Figure~\ref{fig: FD_G}(a). When the E (D) vertex sits at site $q$, 
	the local vertex structure is the same as the case (i) [(ii)].
	Therefore, the forbidden domain is expanded.
	When the G vertices are located at both sites $p$ and $q$,
	the F vertex is located at the site $r$ due to the matching rule of the tiles.
	The forbidden region is shown as the shaded area in Figure~\ref{fig: FD_G}(b).
	Here, we focus on five vertices in the $w$ sublattice outside the shaded area, 
	which is denoted as $q_1, q_2, \cdots, q_5$, to examine their amplitudes.
	According to eq.~(\ref{eq:zero}), the state with $E=0$ is satisfied 
	by the following equations, 
	\begin{eqnarray}
		&&\Psi_{q_1}+\Psi_{q_2}=0,\\
		&&\Psi_{q_2}+\Psi_{q_3}=0,\\
		&&\Psi_{q_1}+\Psi_{q_2}+\Psi_{q_4}=0,\\
		&&\Psi_{q_2}+\Psi_{q_3}+\Psi_{q_5}=0,\\
		&&\Psi_{q_2}+\Psi_{q_4}+\Psi_{q_5}=0.
	\end{eqnarray}
	Thus, we obtain $\Psi_{q_1}=\Psi_{q_2}=\Psi_{q_3}=\Psi_{q_4}=\Psi_{q_5}=0$.
	This means that
	the amplitudes of the vertices on the LH tiles are zero and 
	the forbidden domain can be expanded to the LH tiles.
\end{enumerate}

From these results, we can say that the forbidden domains can be expanded to the whole system
since the S tiles are always isolated in the $\Hzz$ hexagonal golden-mean tiling, 
as shown in Figure~\ref{fig: Deflation_vertex}.
Namely, the amplitude of the wave function is zero in the whole system and
no degenerate states with $E=0$ appear in the $w$ sublattice
under the assumption of the existence of the forbidden domain.
The assumption is equivalent to two conditions $\Psi_i=0$ imposed on the wave function,
where the $i$th site belongs to the $w$ sublattice on a certain P tile.
Therefore, we can prove that the number of the degenerate states with $E=0$ 
in the $w$ sublattice is at most two.

\section{Three groups in the $w$ sublattice} \label{app: B}
Here, we will prove that the $w$ sublattice can be divided into three groups $w_1$, $w_2$ and $w_3$,
so that each site in the $b$ sublattice connects to three nearest-neighbor sites belonging to each of these groups.
Similar to appendix~\ref{app: zero in b}, 
we introduce a domain so that, inside and on its boundary, 
the groups for these vertices are determined.
In the following, we demonstrate that the domain can be expanded to the whole system.
We first focus on a P tile outside of the domain
and adjacent to its boundary, as shown in Figure~\ref{fig: string_P}(a).
When the groups for the sites $p$ and $r$ are determined,
the group of the site $q$ is uniquely determined.
Therefore, the domain can be expanded to include the P tiles according to this rule.

Next, we consider the LH tile outside of the domain and adjacent to its boundary,
as shown in Figure~\ref{fig: string_L}(a).
By considering some cases according to the type of vertex at the site $p$, $z$, etc.,
the group for each site is uniquely and trivially determined
except for two cases, 
which are shown in Figure~\ref{fig: FD_D}(c) and Figure~\ref{fig: FD_G}(b).
In the former (latter) case, the group for the site $p_1$ ($q_2$) is uniquely determined to belong to the groups for the site $r$,
since the two of next nearest-neighbor sites of $p_1$ ($q_2$) belong to the groups for the sites $p, q$.
And the groups for the site $p_2, p_3, \cdots, p_{15}$ ($q_1, q_3, q_4, q_5$) are uniquely determined by the above rules.
Then, we can expand the domain to include the LH tiles.
From these results, we can say that the $w$ sublattice are divided into 
three groups and each site in the $b$ sublattice connects to three nearest-neighbor sites belonging to 
each of these groups.

\bibliography{./refs}

\end{document}